\RequirePackage{lineno}
\documentclass[twocolumn,showpacs,superscriptaddress,amsmath,amssymb]{revtex4-1}
\usepackage{graphicx}
\usepackage{epsfig}
\usepackage{dcolumn}
\usepackage{bm}
\usepackage{overpic}
\usepackage{color}
\usepackage{multirow}
\usepackage{subfigure}

\newcommand{\BR}{{\cal B}}
\newcommand{\R}{{\cal R}}
\newcommand{\eff}{\varepsilon}
\newcommand{\psip}{\psi(2S)}

\newcommand{\jpsi}{J/\psi}
\newcommand{\zcc}{Z_{c}(3900)^\pm}
\newcommand{\zcz}{Z_{c}(3900)^0}
\newcommand{\fz}{f_0(980)}
\newcommand{\ee}{e^+e^-}

\newcommand{\pip}{\pi^+}
\newcommand{\pim}{\pi^-}
\newcommand{\piz}{\pi^0}
\newcommand{\pipi}{\pi^+\pi^-}

\newcommand{\LL}{\ell^+\ell^-}
\newcommand{\lum}{\mathcal{L}}
\newcommand{\ifb}{\rm fb^{-1}}
\newcommand{\ipb}{\rm pb^{-1}}
\newcommand{\gev}{\rm GeV}
\newcommand{\mev}{\rm MeV}
\newcommand{\gevcc}{{\rm GeV}/c^{2}}
\newcommand{\mevcc}{{\rm MeV}/c^{2}}
\newcommand{\eeppjpsi}{\ee\to\piz\piz\jpsi}
\newcommand{\eeppjpsic}{\ee\to\pipi\jpsi}

\begin{document}
\hyphenpenalty=10000
\tolerance=1000

\title{\boldmath Study of the process $e^+e^-\to \pi^0\pi^0 J/\psi$ and neutral charmonium-like state $Z_c(3900)^0$}
\author{
M.~Ablikim$^{1}$, M.~N.~Achasov$^{10,c}$, P.~Adlarson$^{64}$, S. ~Ahmed$^{15}$, M.~Albrecht$^{4}$, A.~Amoroso$^{63A,63C}$, Q.~An$^{60,48}$, ~Anita$^{21}$, Y.~Bai$^{47}$, O.~Bakina$^{29}$
% ...
\\
\vspace{0.2cm}
(BESIII Collaboration)\\
\vspace{0.2cm} {\it
$^{1}$ Institute of High Energy Physics, Beijing 100049, People's Republic of China\\
% ...
$^{4}$ Bochum Ruhr-University, D-44780 Bochum, Germany\\
% ...
$^{10}$ G.I. Budker Institute of Nuclear Physics SB RAS (BINP), Novosibirsk 630090, Russia\\
% ...
$^{15}$ Helmholtz Institute Mainz, Johann-Joachim-Becher-Weg 45, D-55099 Mainz, Germany\\
% ...
$^{21}$ Indian Institute of Technology Madras, Chennai 600036, India\\
% ...
$^{29}$ Joint Institute for Nuclear Research, 141980 Dubna, Moscow region, Russia\\
% ...
$^{47}$ Southeast University, Nanjing 211100, People's Republic of China\\
$^{48}$ State Key Laboratory of Particle Detection and Electronics, Beijing 100049, Hefei 230026, People's Republic of China\\
% ...
$^{60}$ University of Science and Technology of China, Hefei 230026, People's Republic of China\\
% ...
$^{63}$ (A)University of Turin, I-10125, Turin, Italy; (B)University of Eastern Piedmont, I-15121, Alessandria, Italy; (C)INFN, I-10125, Turin, Italy\\
$^{64}$ Uppsala University, Box 516, SE-75120 Uppsala, Sweden\\
% ...
$^{c}$ Also at the Novosibirsk State University, Novosibirsk, 630090, Russia\\
% ...
}
}
 
\date{\today}
\begin{abstract}

  Cross sections of the process $\eeppjpsi$ at center-of-mass energies between 3.808 and 4.600~$\gev$ are measured with high precision by
  using 12.4~$\ifb$ of data samples collected with the BESIII detector operating at the BEPCII collider facility.
A fit to the measured energy-dependent cross sections confirms the existence of the charmonium-like state $Y(4220)$.
  The mass and width of the $Y(4220)$ are determined to be $(4220.4\pm2.4\pm2.3)$~$\mevcc$ and $(46.2\pm4.7\pm2.1)$~$\mev$, respectively, where the first uncertainties are statistical and the second systematic. The mass and width are consistent with those measured in the process $\eeppjpsic$.
  The neutral charmonium-like state $\zcz$ is observed prominently in the $\piz\jpsi$ invariant-mass spectrum, and, for the first time, an amplitude analysis is performed
  to study its properties. The spin-parity of $\zcz$ is determined to be $J^{P}=1^{+}$, and the pole position is $(3893.1\pm2.2\pm3.0)-i(22.2\pm2.6\pm7.0)$~$\mevcc$,
  which is consistent with previous studies of electrically charged $\zcc$.
  In addition, cross sections of $\ee \to \piz\zcz \to\piz\piz\jpsi$ are extracted, and the corresponding line shape is found to agree with that of the $Y(4220)$.
\end{abstract}

\pacs{14.20.Lq, 13.30.Eg, 13.66.Bc, 12.38.Qk}

\maketitle

\section{INTRODUCTION}
A charmonium-like structure with a mass around 4260~$\mevcc$ and spin-parity of $J^{PC}=1^{--}$, namely the $Y(4260)$, was observed and confirmed in the
process $\ee\to (\gamma_{ISR}) \pip\pim\jpsi$ by several experiments~\cite{Babar-Y4260,CLEO-Y4260,2013BelleZc}.
However, the latest results from the BESIII experiment for $\eeppjpsic$ revealed that the $Y(4260)$ consists of two components~\cite{LiuZQ2017},
the $Y(4220)$ and $Y(4320)$. Besides $\eeppjpsic$, dedicated cross-section measurements by BESIII for the processes
$\ee\to\pip\pim h_c$~\cite{BESIII-pipihc}, $\omega\chi_{c0}$~\cite{omegachic0}, $\pipi\psip$~\cite{BESIII-pipipsip},
and $\pip D^0D^{*-}$~\cite{BESIII_piD0Dstar} also support the existence of a structure around 4220~$\mevcc$.
Despite several experimental observations, the internal constituents of the $Y(4220)$ are still unclear.
Possible interpretations include hybrids, meson molecules, hadrocharmonium, or tetraquarks, {\it etc.}, but none of them are
conclusive~\cite{2015MBerwein,2013GuoFK,2014Cleven,2014MBVoloshin,2018Ali,2020Jesse}.

A partial-wave analysis (PWA)~\cite{2017PingRG} showed that the charged charmonium-like state $\zcc$, observed in decays to $\pi^{\pm}\jpsi$
in the process $\eeppjpsic$~\cite{2013BESIIIZc,2013BelleZc}, and to $D^*\bar D$ in the process $\ee\to\pi^{\pm} (D^*\bar D)^{\mp}$~\cite{2014BESIIIZc},
has spin-parity $J^P=1^{+}$.
Different theoretical models~\cite{2015MBerwein,2013GuoFK,2014Cleven,2014MBVoloshin,2018Ali,2020Jesse} consider the $\zcc$ to be an unconventional state; in particular, the molecular and tetra-quark interpretation~\cite{2014Cleven, 2020Jesse}, suggested a strong correlation of the $Y(4220)$ to $\pi Z_c(3900)$.
However, further studies are needed to clarify the situation.

It is particularly interesting to understand the relationship between the $Y$ and $Z_c$ production through the same decay
processes~\cite{BESIII-pipipsip, BESIII_piD0Dstar, 2013BESIIIZc, 2013BelleZc, 2014BESIIIZc}, since such information could shed light on their nature.
More precise measurements of their resonant parameters, production cross sections, and decay modes, as well as searches for new states, are essential.
Studying the neutral decay process $\ee\to\piz\piz\jpsi$ is a natural way to study the properties of the $Y(4220)$ and to search for the charge-neutral
isospin partner of the $\zcc$.

The neutral $\zcz$ was observed in the processes $\ee\to\piz\piz\jpsi$ and $\ee\to\piz(D\bar{D}^*)^0 $ by CLEO-$c$
and BESIII~\cite{2013cleozc, 2015BESIIIZc, BESIII2015_Zcz}, demonstrating that the $\zcc$ is an isovector.
However, the previous analysis were unable to extract the resonant parameters of the $Y$ states and to determine the properties of the $\zcz$,
owing to limited statistics and lack of data in some key center-of-mass (c.m.) energy points.
In this paper, we present an updated analysis of $\ee\to\piz\piz\jpsi$ using 12.4~$\ifb$ of BESIII data taken
at c.m. energies from 3.808 to 4.600~$\gev$.
The production cross sections are measured with high precision, and a PWA is performed to study properties of the $\zcz$.

\section{THE BESIII EXPERIMENT AND THE DATA SETS}

BESIII~\cite{2009MAblikimDet} is a cylindrical spectrometer, consisting of a small-celled, helium-based main drift chamber (MDC), a plastic scintillator time-of-flight system (TOF), a CsI(Tl) electromagnetic calorimeter (EMC), a superconducting solenoid providing a 1.0 T magnetic field, and a muon counter. The charged particle momentum resolution is 0.5\% at a transverse momentum of 1 ${\rm GeV}/c$. The energy loss measurement provided by the MDC has a resolution of 6\%, and the time resolution of the TOF is 80 ps (110 ps) in the barrel (end-caps). The energy resolution for photons is 2.5\% (5\%) at 1 GeV in the barrel (end cap) region of the EMC. 
A more detailed description of the BESIII detector is given in Ref.~\cite{2009MAblikimDet}.

The {\scshape{geant4}}-based~\cite{geant} Monte Carlo (MC) simulation software package {\scshape{boost}}~\cite{boost} is used to determine detection efficiencies and to estimate background rates.
Signal MC samples of $\ee\to\piz\piz\jpsi$ with $\jpsi\to\LL(\ell=e, \mu)$ and $\piz\to\gamma\gamma$, are generated using a phase-space (PHSP) model
and weighted by the amplitude model obtained from the analysis of the data. Potential background contaminations are studied using inclusive MC samples
described in Ref.~\cite{LiuZQ2017}.
Additional exclusive MC samples $\ee\to\eta\jpsi$ with $\eta\to\piz\piz\piz$ and $\ee\to\gamma\psip$ with $\psip\to\piz\piz\jpsi$ are generated to estimate
possible peaking backgrounds.

\section{EVENT SELECTION}

Candidates for charged tracks and photons are selected using the same criteria as described in Ref.~\cite{2015BESIIIZc}.
Electrons (muons) are distinguished with the energy ($E$) deposited in the electromagnetic calorimeter (EMC) divided by their momentum ($p$) measured in the main drift chamber (MDC) with $E/p>0.7$ ($E/p<0.3$).
Candidate $\piz$ decays are reconstructed from pairs of photons with an invariant mass ($M_{\gamma\gamma}$) satisfying $0.11<M_{\gamma\gamma}<0.15~\gevcc$.
Less than three $\piz\piz$ combinations for each event are required to suppress combinatorial backgrounds.
A kinematic fit under the hypothesis $\ee\to\piz\piz\LL$ is applied to enforce energy-momentum conservation and to constrain the two $\piz$ masses.
The combination with the smallest $\chi_{6C}^2$ and with $\chi_{6C}^2<75$ is chosen.

Detailed MC studies indicate that the dominant backgrounds are those from $\ee\to q\bar{q}$, $\ee\to\eta\jpsi$ ($\eta\to\piz\piz\piz$) and
$\gamma\psip$ ($\psip\to\piz\piz\jpsi$). The first one has a uniform distribution in the $\LL$ invariant-mass ($M_{\LL}$) spectrum,
while the other two produce a broad peak around the $\jpsi$ position.
To determine the signal yield of $\ee\to\piz\piz\jpsi$, an unbinned maximum-likelihood fit to the $M_{\LL}$ spectrum in the range $2.90<M_{\LL}<3.30~\gevcc$ is performed.
The signal and peaking backgrounds are described using MC-simulated shapes convolved with a common Gaussian function representing the resolution difference between
data and the MC simulation, while the other possible backgrounds are described with a first-order polynomial function.
The peaking background contributions are normalized to the integrated luminosity.

\section{STUDY OF THE $\ee\to\piz\piz\jpsi$ LINE SHAPE}
\subsection{EXTRACTION OF THE BORN CROSS SECTION}
The Born cross section is determined by
\begin{equation}
\sigma^{\rm B}=\frac{N^{\rm obs}}{\lum_{\rm int}~(1+\delta^{r})~(1+\delta^{v})~\eff~\BR},
\end{equation}
where $N^{\rm obs}$ is the signal yield obtained as described above, $\lum_{\rm int}$ is the integrated luminosity,
$\eff$ is the selection efficiency estimated with signal MC samples modeled according to the results of the PWA for those points that fall within
$\sqrt{s}=4.178\sim4.416~\gev$ and by PHSP for other energy points with low statistics,
$(1+\delta^{r})$ is the radiative correction factor, $(1+\delta^{v})$ is the vacuum polarization factor derived from QED calculations~\cite{vacuum} and
$\BR$ is the world averaged value~\cite{2019PDG} of the product branching fractions (BFs) of the intermediate states.
The resultant cross sections are shown in the top subplot of Fig.~\ref{fig:lineshape}, and the various numbers used in the calculation are summarized in Table~\ref{tab:crs}. 
To examine the isospin symmetry, the cross-section ratios of the processes $\ee\to\piz\piz\jpsi$ to $\ee\to\pipi\jpsi$~\cite{LiuZQ2017}
at different c.m. energies are extracted, and are fitted by a constant function, as shown in the bottom subplot of Fig.~\ref{fig:lineshape}.
The fit yields an average ratio of $0.48\pm0.02$, consistent with the prediction 0.5 based on isospin symmetry.

 \begin{table*}[htbp]
  \begin{center}
  \small
  \caption{Summary of the center-of-mass energies ($\sqrt{s}$), luminosities ($\lum$), detection efficiencies estimated from MC samples generated by PWA resulted amplitudes ($\eff^{\rm PWA}$) and from PHSP MC samples ($\eff^{\rm PHSP}$), numbers of estimated peaking backgrounds $\ee\to\eta\jpsi$ ($N^{\rm est}_{\eta\jpsi}$), $\ee\to\gamma\psip$ ($N^{\rm est}_{\gamma\psip}$) and observed signal events ($N_{\piz\piz\jpsi}^{\rm obs}$), radiative correction factors (1+$\delta^{r}$), vacuum polarization factors (1+$\delta^{v}$), Born cross sections of $\ee\to\piz\piz\jpsi$ ($\sigma^{\rm Born}$) and the cross section ratios of $\ee\to\piz\piz\jpsi$ to $\ee\to\pip\pim\jpsi$, where the first uncertainties are statistic and second systematic. The ratios at some of these energy points are not available due to the lack of published cross section of $\ee\to\pip\pim\jpsi$.
}
  \begin{tabular}{c c c c c c c c c c c}
  \hline\hline
  $\sqrt{s}(\gev)$  &$\lum({\rm \ipb)}$  & $\eff^{\rm PWA}(\%)$ & $\eff^{\rm PHSP}(\%)$ &$N^{\rm est}_{\gamma\psip}$ &$N^{\rm est}_{\eta\jpsi}$  &$N_{\piz\piz\jpsi}^{\rm obs}$  & 1+$\delta^{r}$  &1+$\delta^{v}$  & $\sigma^{\rm Born}$ (pb) &$\R(\frac{\piz\piz\jpsi}{\pi^{+}\pi^{-}\jpsi})$\\ \hline

  3.808 &  $50.5\pm0.5$  &    & 20.44 &  &  & $11.5\pm4.0$  &~0.815~ &~1.056~   &~$11.11\pm 3.87 \pm 0.78$~ &~$0.70\pm0.28\pm0.17$~\\
  3.896 &  $52.6\pm0.5$  &    & 19.89 &  & &  $6.5\pm3.3$  & 0.855 & 1.049     & $5.95 \pm 3.02 \pm 0.42$  & $0.36\pm0.23\pm0.10$ \\
  4.008 & $482.0\pm4.8$  &  & 19.64 &11.3 & 1.4 & $88.6\pm11.2$ & 0.904 & 1.044     & $8.51 \pm 1.14 \pm 0.60$  & $0.56\pm0.08\pm0.07$ \\
  4.086 &  $52.9\pm0.4$  &    & 19.60 &  &  & $10.6\pm3.6$  & 0.933 & 1.052     & $8.96 \pm 3.04 \pm 0.63$  & $0.63\pm0.22\pm0.14$ \\
  4.178 &$3194.5\pm31.9$ & 20.68 & 20.09 &25.4 &8.6  &$365.7\pm24.6$ & 0.900 & 1.054     & $5.00 \pm 0.34 \pm 0.35$  &     \\
  4.189 &  $43.3\pm0.3$  &    & 20.97 &  &  & $11.2\pm3.8$  & 0.861 & 1.056     & $11.65\pm 3.91 \pm 0.82$  & $0.79\pm0.27\pm0.21$ \\
  4.189 & $524.6\pm0.4$  & 22.91 & 20.78 & 4.4 &2.4  & $89.3\pm11.4$ & 0.859 & 1.056     & $7.03 \pm 0.90 \pm 0.49$  & $0.48\pm0.06\pm0.12$ \\
  4.200 & $526.0\pm0.4$  & 22.01 & 21.54 &4.6  &2.6  &$122.6\pm12.5$ & 0.810 & 1.056     & $10.62\pm 1.08 \pm 0.74$  &           \\
  4.208 &  $55.0\pm0.4$  &   & 22.44 &  & & $26.9\pm5.7$  & 0.768 & 1.057     & $23.10\pm 4.86 \pm 1.62$  & $0.46\pm0.10\pm0.06$ \\
  4.210 & $518.0\pm0.4$  & 21.94 & 22.16 &4.5  &2.9  &$235.9\pm16.7$ & 0.762 & 1.057     & $22.12\pm 1.57 \pm 1.55$  & $0.44\pm0.03\pm0.05$ \\
  4.217 &  $54.6\pm0.4$  &    & 22.89 &  &  & $28.6\pm5.8$  & 0.741 & 1.057     & $25.07\pm 5.08 \pm 1.76$  & $0.44\pm0.09\pm0.06$ \\
  4.219 & $514.6\pm0.4$  & 22.95 & 22.60 &4.3  &3.0  &$351.0\pm20.1$ & 0.739 & 1.056     & $32.66\pm 1.87 \pm 2.29$  & $0.57\pm0.03\pm0.07$  \\
  4.226 &$1100.9\pm7.0$  & 22.75 & 23.27 &8.8  &5.2  &$890.4\pm32.1$ & 0.744 & 1.056     & $38.84\pm 1.40 \pm 2.72$  & $0.48\pm0.02\pm0.04$ \\
  4.236 & $530.3\pm0.5$  & 21.80 & 23.22 &4.4  &2.4  &$452.9\pm22.6$ & 0.779 & 1.056     & $40.88\pm 2.04 \pm 2.86$  &              \\
  4.242 &  $55.9\pm0.4$  &    & 23.26 &  & & $49.1\pm7.6$  & 0.805 & 1.056     & $38.15\pm 5.87 \pm 2.67$  & $0.48\pm0.07\pm0.06$ \\
  4.244 & $538.1\pm0.5$  & 23.34 & 23.04 &3.9  &1.9  &$435.3\pm22.0$ & 0.815 & 1.056     & $34.59\pm 1.75 \pm 2.42$  &          \\
  4.258 & $828.4\pm5.5$  & 22.91 & 23.15 &5.9  &1.3  &$585.1\pm25.9$ & 0.857 & 1.054     & $29.29\pm 1.30 \pm 2.05$  & $0.52\pm0.02\pm0.05$ \\
  4.267 & $531.1\pm0.6$  & 24.64 & 22.68 &3.6  &0.8  &$356.0\pm20.6$ & 0.869 & 1.053     & $25.51\pm 1.48 \pm 1.79$  &              \\
  4.278 & $175.7\pm0.6$  & 22.98 & 22.44 &1.1  &0.2  &$110.0\pm11.2$ & 0.877 & 1.053     & $25.32\pm 2.58 \pm 1.77$  &             \\
  4.308 &  $45.1\pm0.3$  &    & 22.35 &  & & $23.3\pm5.4$  & 0.886 & 1.052     & $21.32\pm 4.91 \pm 1.49$  & $0.43\pm0.10\pm0.06$ \\
  4.358 & $543.9\pm3.6$  & 21.24 & 20.38 &2.9  &0.4  &$196.4\pm15.9$ & 1.062 & 1.051     & $13.07\pm 1.06 \pm 0.92$  & $0.54\pm0.04\pm0.06$ \\
  4.387 &  $55.6\pm0.4$  &  & 18.52 &  &  &  $4.2\pm2.6$  & 1.177 & 1.051     & $2.83 \pm 1.77 \pm 0.20$  & $0.15\pm0.09\pm0.03$ \\
  4.416 &$1090.7\pm6.9$  & 17.59 & 17.18 &4.7  &1.1  &$182.8\pm16.2$ & 1.238 & 1.052     & $6.27 \pm 0.56 \pm 0.44$  & $0.55\pm0.05\pm0.06$ \\
  4.467 & $111.1\pm0.7$  &    & 16.28 & &  & $17.2\pm4.9$  & 1.268 & 1.055     & $6.11 \pm 1.72 \pm 0.43$  & $0.48\pm0.14\pm0.09$ \\
  4.527 & $112.1\pm0.7$  &    & 16.32 &  &  & $14.1\pm5.0$  & 1.265 & 1.055     & $4.96 \pm 1.74 \pm 0.35$  & $0.49\pm0.17\pm0.10$ \\
  4.575 &  $48.9\pm0.1$  &    & 16.52 &  & &  $7.2\pm3.1$  & 1.259 & 1.055     & $5.72 \pm 2.47 \pm 0.40$  & $0.45\pm0.19\pm0.12$ \\
  4.600 & $586.9\pm3.9$  &    & 16.76 &1.8  &0.1  & $32.6\pm7.8$  & 1.252 & 1.055     & $2.15 \pm 0.52 \pm 0.15$  & $0.36\pm0.09\pm0.05$ \\ \hline

  \hline\hline
  \end{tabular}
  \label{tab:crs}
  \end{center}
  \end{table*}

 \begin{figure}[htbp]
  \centering
  \mbox{
     \includegraphics[width=0.4\textwidth, height=0.3\textwidth]{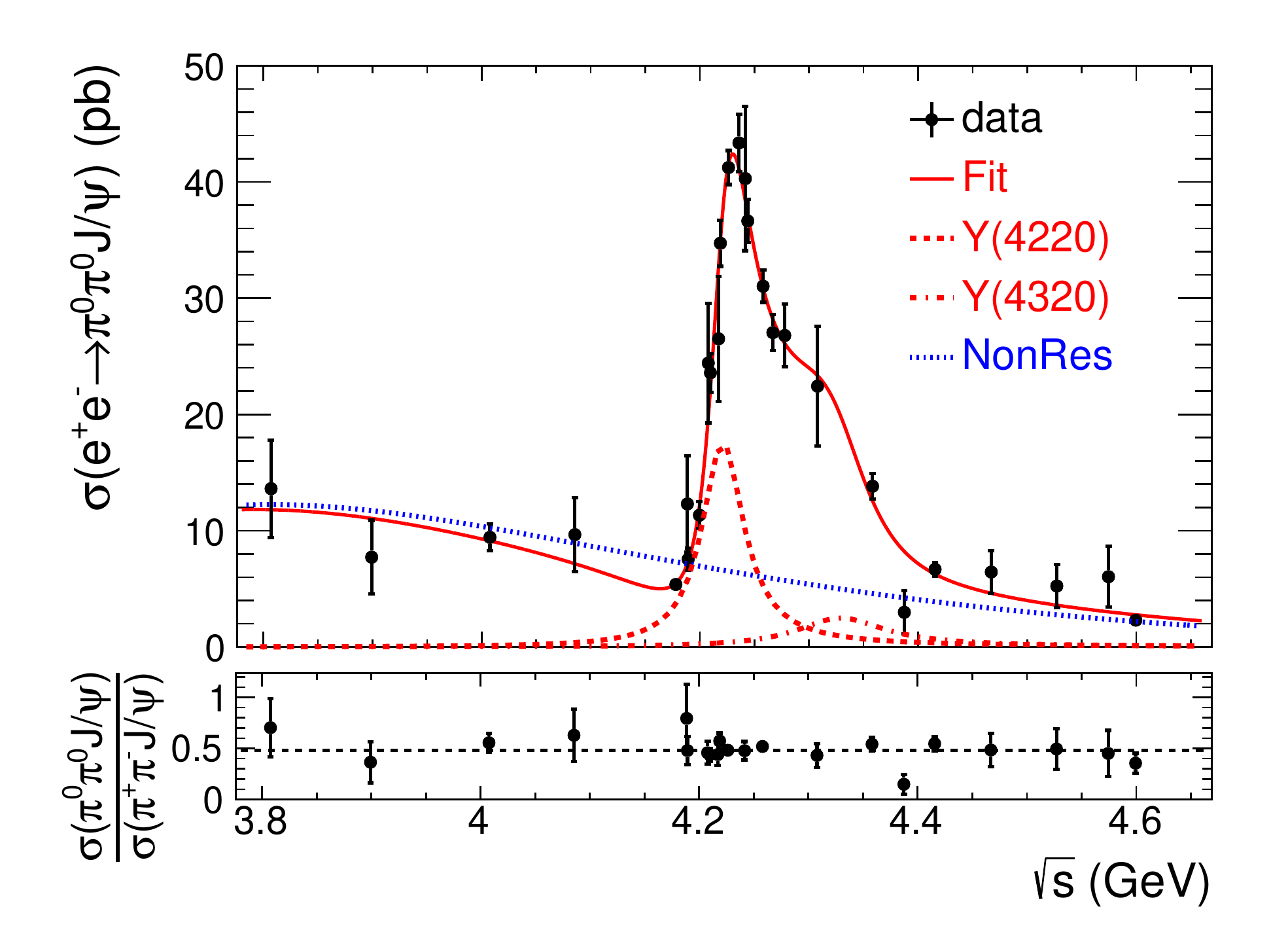}
  }
  \caption{(Color online) Top: Fit to the Born cross sections of $\ee\to\piz\piz\jpsi$, where points with error bars are data, the red solid line is the
    total fit result, the blue dotted line is the non-resonant component, while the red dashed and dot-dashed lines represent the contributions
    from $Y(4220)$ and $Y(4320)$, respectively. Bottom: Cross-section ratio of $\ee\to\piz\piz\jpsi$ to $\ee\to\pip\pim\jpsi$, where
    the black dashed line corresponds to the average.}
\label{fig:lineshape}
\end{figure}

\subsection{FIT TO THE CROSS SECTION}
A $\chi^{2}$ fit is performed to the Born cross section to study resonant structure in the range $\sqrt{s}=3.808\sim4.600~\gev$.
The cross section line shape is described by
\begin{equation}
\sigma_{\rm fit}(\sqrt{s})=|\sqrt{\sigma_{NY}(\sqrt{s})}+f_{1}(s)e^{i\phi_{1}}+f_{2}(s)e^{i\phi_{2}}|^{2},
\label{eq:expo}
\end{equation}
where $\sigma_{NY}(\sqrt{s})=\Phi(\sqrt{s})e^{-p_{0}(\sqrt{s}-M_{\rm thd})+p_{1}}$ represents the non-resonant component~\cite{Babar-Y4008},
$M_{\rm thd}=2m_{\piz}+m_{\jpsi}$,
$\Phi(\sqrt{s})$ 
is the PHSP factor of the three-body decay $R_{i}\to\piz\piz\jpsi$~\cite{2019PDG}, 
$f_i$ stands for the amplitude of structure $R_i$ ($R_1$ is $Y(4220)$ and $R_2$ is $Y(4320)$), and $\phi_i$ is its phase relative to the continuum. The amplitude $f_i$ is defined as
\begin{equation}
f_i(s)=\frac{M_i}{\sqrt{s}}\frac{\sqrt{12\pi\Gamma^i_{ee}\Gamma^i_{tot}\BR^i_{\piz\piz\jpsi}}}{s-M_i^{2}+iM_i\Gamma^i_{tot}}\times\sqrt{\frac{\Phi(\sqrt{s})}{\Phi(M_i)}},
\end{equation}
where $M_i$, $\Gamma^i_{tot}$, $\Gamma^i_{ee}$ and $\BR^i_{\piz\piz\jpsi}$ are the mass, full width, partial width coupling to $\ee$,
and the decay BF of $R_i\to\piz\piz\jpsi$, respectively.
The mass and width of the $Y(4320)$ are fixed to those reported in Ref.~\cite{LiuZQ2017} and other parameters are allowed to vary.
There are four solutions owing to degeneracies from the phase $\phi(R_1)$ of the dominant component $Y(4220)$: the difference in phase between solutions I and II (III and IV) is approximately 90 degrees, while solutions III and I (IV and II) differ by around 45 degrees. A summary of the fit results can be found in Table~\ref{tab:fit2R_pars}. 
The goodness of fit is $\chi^{2}/{\rm ndf}=19.3/19$, where ${\rm ndf}$ is the number of degrees of freedom. The mass and width of the $Y(4220)$ are measured to be $(4220.4\pm2.4)~\mevcc$ and $(46.2\pm4.7)~\mev$, respectively, which agree with the ones obtained in $\eeppjpsic$~\cite{LiuZQ2017}.
The fit curves for one of the solutions are shown in Fig.~\ref{fig:lineshape}. 
The statistical significance of $Y(4320)$ is estimated to be 4.2$\sigma$ by the changes in the $\chi^2$ and ${\rm ndf}$ values obtained from including and excluding
the $Y(4320)$.

\begin{table*}[htbp]
\begin{center}
\begin{small}
\caption{Summary of the fit results to the measured cross sections of $\ee\to\piz\piz\jpsi$. The uncertainties are statistical only.}
\begin{tabular}{c c c c c}\hline\hline
Parameters  & ~~Solution I~~  & ~~Solution II~~  & ~~Solution III~~  & ~~Solution IV~~  \\\hline
$p_{0} (c^2/{\rm MeV})$			 &\multicolumn{4}{c}{ 5.1$\pm$ 0.5}   \\
$p_{1}$			 &\multicolumn{4}{c}{ (3.0$\pm$ 2.9)$\times 10^{-2}$}   \\
$M({R_1})~(\mevcc)$ 			 	  &\multicolumn{4}{c}{ 4220.4$\pm$2.4 }   \\
$\Gamma_{\rm tot}(R_1)~(\mev)$        &\multicolumn{4}{c}{ 46.2$\pm$4.7}   \\
$\Gamma^1_{\rm ee}\BR^1_{R_{1}\to\piz\piz\jpsi}~(\rm eV)$   &0.99$\pm$0.17 & 4.13$\pm$0.28  &1.38$\pm$0.30  &5.72$\pm$0.57    \\
$\Gamma^2_{\rm ee}\BR^2_{R_2\to\piz\piz\jpsi}~(\rm eV)$   &0.31$\pm$0.15 & 0.41$\pm$0.20  &6.44$\pm$0.41  &8.51$\pm$0.44    \\
$\phi(R_1)$ 		&$(16.8\pm5.5)^\circ$      &$(-73.6\pm3.6)^\circ$   &$(63.0\pm8.4)^\circ$  &$(-28.2\pm2.0)^\circ$   \\ 
 $\phi(R_2)$ 		& $(133.5\pm12.3)^\circ$   &$(99.2\pm1.6)^\circ$   &$(-96.0\pm5.2)^\circ$   &$(-131.3\pm2.8)^\circ$    \\
\hline\hline
\end{tabular}
\label{tab:fit2R_pars}
\end{small}
\end{center}
\end{table*}

\subsection{SYSTEMATIC UNCERTAINTIES}
The systematic uncertainties for the cross section measurement include those associated with the luminosity, detection efficiency, radiative correction,
fitting procedure, and the BFs of intermediate states.
The uncertainties of the detection efficiency include those of tracking, photon detection, $e/\mu$ identification, $\piz$ reconstruction, kinematic fit, and MC modeling.
The uncertainty of the tracking efficiency for leptons is $1.0\%$ per track~\cite{systrack}.
The uncertainty for the photon detection is 1.0\% per photon~\cite{sysphoton}.
The uncertainty of the $e/\mu$ identification is 0.2\% per track estimated with a control sample $\ee\to\piz\piz\jpsi$ at $\sqrt{s}=3.686~\gev$.
The uncertainty for $\piz$ reconstruction, 0.3\%, is the efficiency difference by smearing the $M_{\gamma\gamma}$ distribution of signal MC sample to match that of data.
The uncertainty due to the kinematic fit is 2.6\%, estimated with a clean control sample $\ee\to\gamma\psip,~\psip\to\piz\piz\jpsi$ at $\sqrt{s}=4.178\gev$.
The uncertainty associated with the MC modeling is estimated by two approaches.
For high-statistics c.m. energy points for which PWA is performed,
we generate 100 sets of signal MC samples at each c.m. energy point  by altering the input parameters with an additional residual generated randomly with a multi-variable Gaussian function according to the PWA results,
and the resultant standard deviation of detection efficiencies is taken as the systematic uncertainty.
We also take differences between the nominal efficiencies and efficiencies estimated by PHSP MC samples at the c.m. energy points with large statistics as the uncertainty for those with low statistics.
The uncertainty from initial and final state radiation depends on both the radiative correction factor uncertainty of 0.3\%, obtained by altering the input cross section line shape from Ref.~\cite{LiuZQ2017} to the measurements in this analysis, as well as the vacuum polarization factor uncertainty of 0.5\% taken from a QED calculation~\cite{vacuum}. 
The uncertainty of the fit procedure, 1.2\%, including uncertainties of the background shape and the fit range, is estimated by altering the first order polynomial function to the quadratic and by varying the fit range by 100~$\mevcc$; 
the (largest) change of signal yield is taken as the uncertainty.
The uncertainty of peaking backgrounds is negligible by considering the uncertainties of their cross sections.
The uncertainties on the BFs of $\jpsi\to\LL$ and $\piz\to\gamma\gamma$ are quoted from Ref.~\cite{2019PDG}.
All of the above uncertainties are summarized in Table~\ref{tab:sys_crs}; assuming all sources are independent, the total uncertainty is found to be 6.1\% by adding all individual values in quadrature.
  \begin{table}[htbp]
  \begin{center}
  \small
  \caption{Summary of the systematic uncertainties of the $\ee\to\piz\piz\jpsi$ cross section measurement. }
  \begin{tabular}{l c  }
      \hline \hline
      Sources 	& Uncertainties ($\%$)  \\ \hline
      Tracking for $e^{\pm}/\mu^{\pm}$   & 2.0           \\
      E/p requirement  		          & 0.4            \\
      Photon efficiency 		  & 4.0            \\
      $\piz$ mass window 		  & 0.3            \\
      Kinematic fit               & 2.6       \\
      Fit to $M_{\LL}$            & 1.2           \\
      Radiative correction        & 0.3       \\
     Vacuum polarization       & 0.5       \\
      MC model         		      & 2.7         \\
      Luminosities                & 1.0       \\
      $\BR_{\rm{inter}}$          & 0.6      \\  \hline
      Total                       & 6.1           \\
      \hline\hline
  \end{tabular}
  \label{tab:sys_crs}
  \end{center}
  \end{table}

The uncertainties for the $Y(4220)$ resonant parameters are from the measured cross section and the fit procedure, as well as the c.m.~energy measurement and its spread.
The c.m.~energy uncertainty  0.8~$\mev$, obtained from a measurement of di-muon events~\cite{2016Ecms}, is directly propagated to the mass of the $Y(4220)$.
The uncertainties from the c.m.~energy spread are obtained by convolving the resonant PDF with a Gaussian whose width is taken to be 1.6~$\mev$, equal to the spread obtained from the Beam Energy Measurement System~\cite{enespread}.
The uncertainties associated with the fit procedure include those of the PDF model, the PHSP factor, and the resonant parameters of the $Y(4320)$.
The uncertainties of the PDF model are estimated by replacing the exponential function with a broad resonance $Y(4008)$~\cite{LiuZQ2017} and by changing the constant full width $\Gamma_{\rm tot}$ to a phase-space-dependent full width $\Gamma_{\rm tot}\frac{\Phi(\sqrt{s})}{\Phi(M)}$.
The uncertainty of the three-body PHSP factor, due to the existence of intermediate states, is
estimated by considering the PHSP of cascade two-body decays of $\ee\to R\jpsi$ with $R\to\piz\piz$ ($R=\sigma,~f_0(980), ~f_0(1370)$) and $\ee\to\piz\zcz$ with $\zcz\to\piz\jpsi$.
The uncertainties associated with the resonant parameters of the $Y(4320)$ are obtained by changing their values
by 1$\sigma$ of their uncertainty.
In the above scenarios, the alternative fits are performed, and the resultant differences with respect to the nominal values are considered as the uncertainties.
As summarized in Table~\ref{tab:sys4220}, assuming all of the systematic uncertainties are independent, the total uncertainties of 2.3~$\mevcc$ and 2.1~$\mev$ for the mass and width of $Y(4220)$, respectively, are the quadratic sum of the individual values.

  \begin{table}[htbp]
  \begin{center}
  \small
  \caption{Summary of the systematic uncertainties of the $Y(4220)$ resonant parameters. }
  \begin{tabular}{ l  c c }
  \hline \hline
  \multirow{2}{*}{Sources}  & \multicolumn{2}{c}{Uncertainties}  \\ \cline{2-3}
                & ~~~~M~$(\mevcc)$~~~~  &~~~~$\Gamma$~$(\mev)$~~~~  \\ \hline
      Cross section measurement &0.2   & 0.3          \\
      c.m. energies    &0.8   &  $-$         \\
      c.m. energy spread  &0.01   & 0.4          \\
      Fit procedure &2.2   & 2.0     \\ \hline
      Total        &2.3   & 2.1           \\
  \hline\hline
  \end{tabular}
  \label{tab:sys4220}
  \end{center}
  \end{table}

\section{STUDY OF THE NEUTRAL $\zcz$ STATE}

A PWA is applied to measure the spin and parity of $\zcz$. The candidate signal events are required to be within the $\jpsi$ signal region ($3.06<M_{\LL}<3.13$~$\gevcc$).
Events within the $\jpsi$-sideband region $((2.93, 3.00)\cup(3.20, 3.27))~\gevcc$ are used to evaluate the effect of non-$\jpsi$ background.
The $\jpsi$-peaking background is found to be negligible.
To improve the purity of samples in the $\jpsi\to\mu^+\mu^-$ channel, at least one muon track penetrating more than five muon chamber (MUC) layers is required.
The invariant-mass distributions of $\piz\jpsi$ ($M_{\piz\jpsi}$) and $\piz\piz$ ($M_{\piz\piz}$) for the four data samples of $\sqrt{s}=$~4.226, 4.236, 4.244 and 4.258~$\gev$, which have the largest statistics among all data samples, are shown in Fig.~\ref{fig:invmass}, where the structures $\zcz$ and $f_0(980)$ are clearly visible.
To decompose the intermediate processes, a simultaneous PWA is performed on the four data samples with a sum of $2188 \pm 47$ candidate events, of which $57 \pm 8$ are background events estimated from the $\jpsi$-sideband regions. 

 \begin{figure*}[htbp]
  \centering
  \mbox{
    \begin{overpic}[width=0.32\textwidth, height=0.95\textwidth]{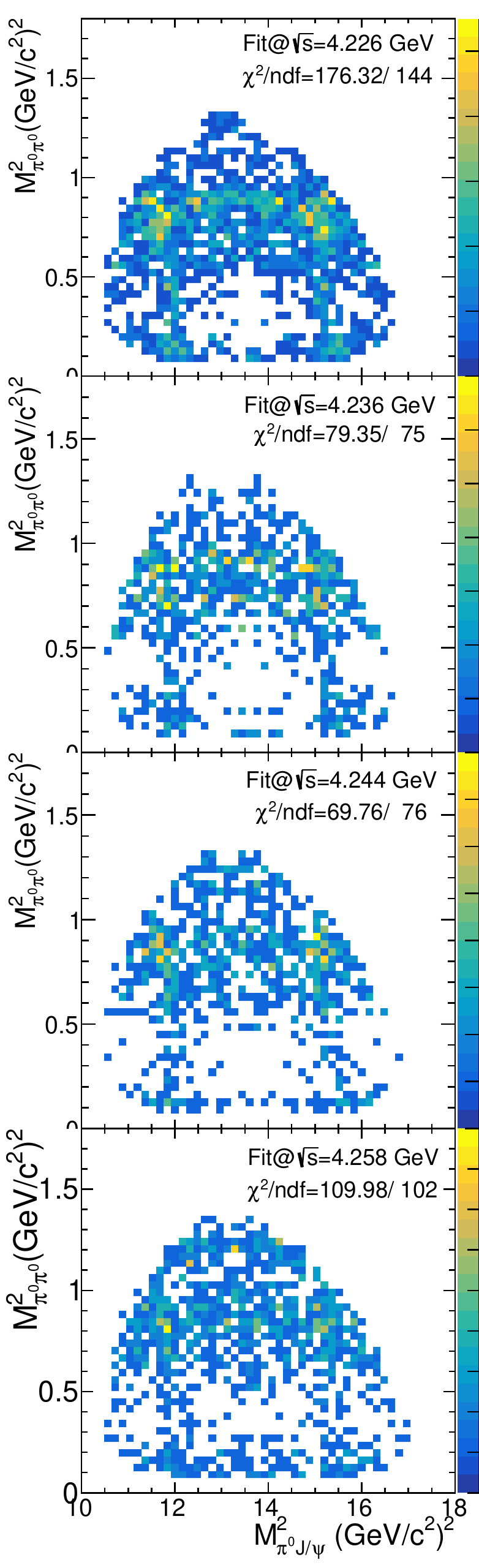}
    \end{overpic}
    \begin{overpic}[width=0.67\textwidth, height=0.95\textwidth]{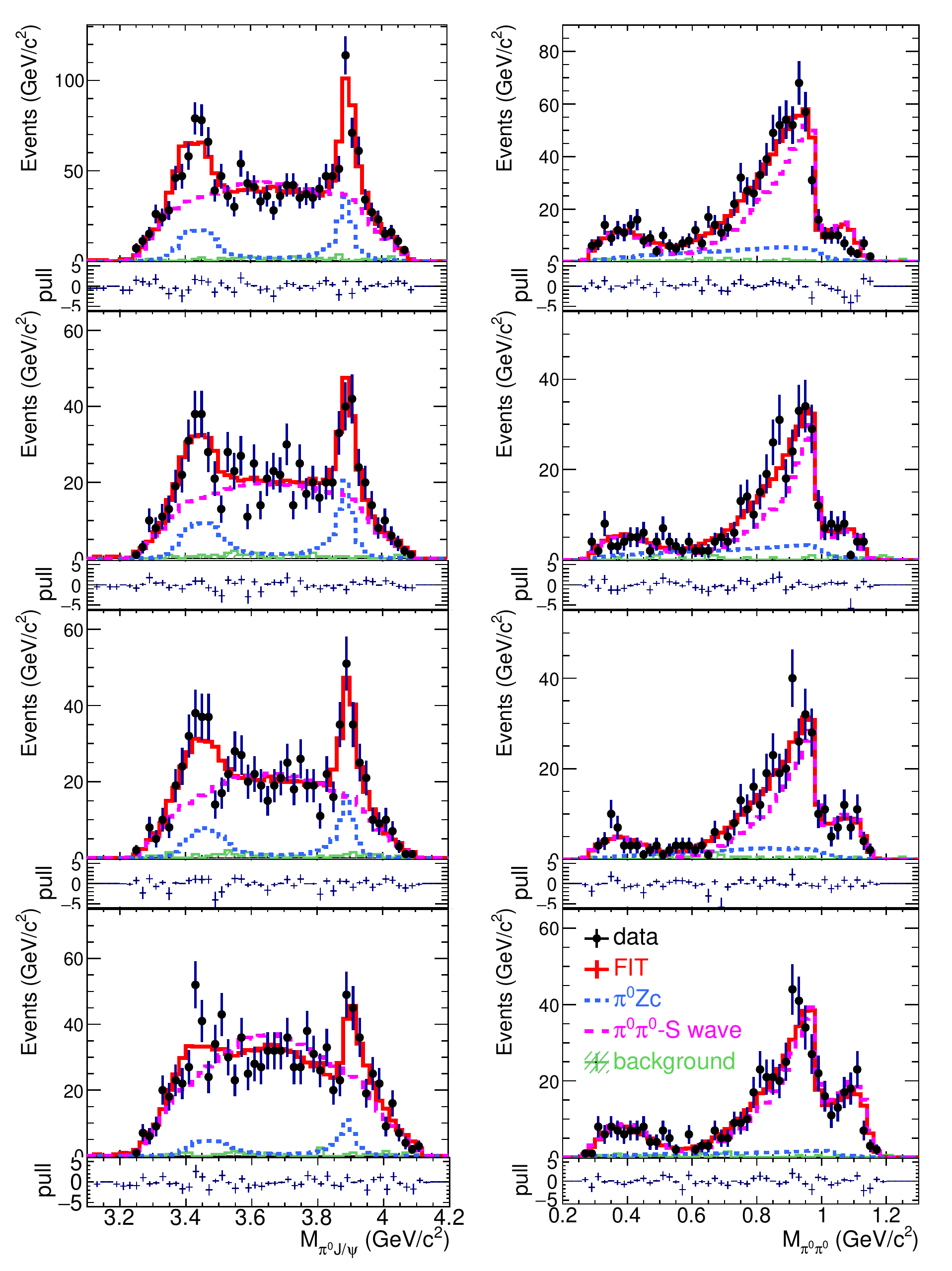}
    \end{overpic}
  }
  \caption{(Color online) (Left column) Dalitz plots of $M^2_{\piz\jpsi}$ versus $M^2_{\piz\piz}$, invariant-mass projections (middle column) $M_{\piz \jpsi}$ and (right column) $M_{\piz\piz}$ of the results of the nominal PWA for data samples $\sqrt{s}=4.226\sim4.258~\gev$. Points with errors are data, red solid curves are the total fit results, the blue dashed (magenta long-dashed) curves represent $\zcz$ ($\piz\piz$-S wave) components, and green shaded histograms represent the estimated backgrounds. Each event appears twice in the Dalitz plots and $M_{\piz\jpsi}$ distributions. The $\chi^2/{\rm ndf}$ is calculated by merging those bins with less than 10 events in the Dalitz plots.}
  \label{fig:invmass}
\end{figure*}

\subsection{AMPLITUDE CONSTRUCTION}
 The isobar model is implemented in the PWA and includes the cascading decay chains $\ee\to \piz\zcz\to\piz\piz\jpsi$ and $\ee\to R\jpsi \to\piz\piz\jpsi$,
 where $R$ represents $\sigma$, $\fz$, $f_0(1370)$, {\it etc.}.
The quasi two-body decay amplitudes in the sequential decays are constructed using the helicity formalism~\cite{helicity}.  For a decay of particle A, $A(J, M)\to 1(J_1, \lambda_1)+2(J_2, \lambda_2)$, where spin and helicity are indicated in the parentheses, the amplitude is given by
\begin{equation}
A_{\lambda_{1}, \lambda_{2}}(\theta, \phi) = N_{J}F^{J}_{\lambda_{1}, \lambda_{2}}D^{*J}_{M, \lambda}(\phi, \theta, 0),~ (\lambda=\lambda_{1}-\lambda_{2}), 
\end{equation}
where $N_{J}$ is the normalization factor, $F^{J}_{\lambda_{1}, \lambda_{2}}$ is the helicity-coupling amplitude constrained by the parity conservation and $D^{J}_{M, \lambda}(\phi, \theta, 0)$ is to describe the angular distribution of the final state particle with its polar ($\theta$) and azimuthal ($\phi$) angles in the rest frame of the mother particle.
$F^{J}_{\lambda_{1}, \lambda_{2}}$ is cited from Chung's formula~\cite{helicity}:
\begin{equation}
\begin{aligned}
F^{J}_{\lambda_{1}, \lambda_{2}}=&\sum_{LS}g_{LS}^{J}\sqrt{\frac{2L+1}{2J+1}}   ~~\\
&<L0S\lambda|J\lambda><S_{1}\lambda_{1}S_{2}-\lambda_{2}|S\lambda>p^{L}B_{L}(p, r),
\end{aligned}
\label{Eq:Ffunc}
\end{equation}
where $g_{LS}$ is the coupling constants in the $LS$ coupling scheme, the angular brackets denote Clebsch-Gordan coefficients, and the orbital angular momentum barrier factor $p^{L}B_{L}(p, r)$ involves the Blatt-Weisskopf functions~\cite{helicity}. 

The total amplitude is a sum of the amplitudes of two different cascading decay chains, 
\begin{widetext}
\begin{eqnarray}
\begin{aligned}
I=\sum_{\lambda_{Y},\Delta\lambda_{l}}\mid \sum_{\lambda_{Z_c^0},\lambda_{R_j},\lambda_{\jpsi}}&[\sum_jA^{Y\to R_j\jpsi}_{\lambda_{R_j}, \lambda_{\jpsi}}(\theta_{R_j}, \phi_{R_j})BW(M_{\piz\piz})A^{R_j\to\piz\piz}_{0,0}(\theta_{\piz}, \phi_{\piz})A^{\jpsi\to\LL}_{\lambda_{l^+}, \lambda_{l^-}}(\theta_{l^+}, \phi_{l^+}) ~\\
 &+ e^{i\Delta\lambda_{l}\alpha_{l}}A^{Y\to\piz Z_c^0}_{\lambda_{Z_c^0},0}(\theta_{Z_c
^0}, \phi_{Z_c^0})BW(M_{\piz\jpsi})A^{Z_c^0\to\piz\jpsi}_{\lambda_{\jpsi},\lambda_{\piz}}(\theta_{\jpsi}, \phi_{\jpsi})A^{\jpsi\to\LL}_{\lambda_{l^+}, \lambda_{l^-}}(\theta_{l^+}, \phi_{l^-})] \mid ^2
\end{aligned}
\label{eq:totamp}
\end{eqnarray}
\end{widetext}
where $\lambda_{P}$ is the helicity of particle $P$, ($\theta_{P}$, $\phi_P$) are the polar and azimuthal angles of particle $P$ in the helicity frame of the cascading decay, $M_{\piz\piz}$ and $M_{\piz\jpsi}$ are the invariant mass of $\piz\piz$ and $\piz\jpsi$, respectively.
The amplitude of $\jpsi\to\LL$ is considered, and an additional term $e^{i\Delta\lambda_{l}\alpha_{l}}$ is introduced to correct for the difference in the azimuthal angle ($\alpha_l$) between the lepton helicities in two different decay chains~\cite{Belle2013_LHCb2015}.
The intermediate states are parameterized with the relativistic Breit-Wigner (BW) functions, except for the $\fz$, which is described by a Flatt$\acute{e}$ formula with
fixed parameters taken from Ref.~\cite{f0980flatte}. 
The width of the wide $\sigma$ resonance is parameterized as $\sqrt{1-\frac{4m^2_{\pi}}{s}}\Gamma$~\cite{E791sigma}.
The resonant parameters of the $f_0(1370)$ and other well-known mesons are taken from the world averaged values~\cite{2019PDG},
while those of the $\zcz$ are left free in the fit.
The relative magnitudes and phases of the individual intermediate processes are determined by performing an unbinned extended-maximum-likelihood fit
for the four data samples simultaneously.
The overall likelihood is given by:
\begin{equation} 
L=\frac{e^{-\mu}\mu^{N}}{N!}\prod^{N}_{i=1}\frac{I(\Omega_{i}, \alpha)\eta(\Omega_{i})}{\mu} 
\end{equation}
where $I(\Omega_{i}, \alpha)$ is the total amplitude squared as defined in Eq.(\ref{eq:totamp}) with a set of of four-vector momenta $\Omega_i=(p_{\piz}, p_{\piz}, p_{\ell^{+}}, p_{\ell^{-}})_i$ and float parameters $\alpha$; $\eta(\Omega)$ is the signal selection efficiency; $N$ is the observed number of events and $\mu$ is the normalization factor. The normalization factor $\mu=\int{I(\Omega, \alpha)\eta(\Omega)d\Omega}$ is calculated as the sum of $I(\Omega, \alpha)$ over $\ee\to\piz\piz\jpsi$ events generated with the PHSP model that passed through the detector simulation and event selection.   
The background contribution to the overall likelihood value is estimated from the events in the $\jpsi$-sideband region and subtracted.
The properties of the intermediate states, namely their masses, widths and coupling strengths to the different final states are shared
parameters between the different data samples, while the production magnitudes and the relative phases are independent.
All of the possible known intermediate states are considered in the fit and only those with statistical significances larger than 5$\sigma$ are kept in the nominal results.
The statistical significance is calculated by the differences in the likelihood values and the number of degrees of freedom between the two scenarios with or without the intermediate process included. Different $\zcz$ spin-parity hypotheses are tested in the simultaneous fit.

\subsection{PWA RESULTS}
With the above strategy, the nominal fit includes the intermediate processes $\ee\to\sigma \jpsi$, $\fz\jpsi$, $f_0(1370)\jpsi $ and $\piz\zcz$.
As shown in Fig.~\ref{fig:invmass}, the $\piz\piz$ S-wave contribution dominates. 
The spin-party of the $\zcz$ is determined to be $J^P=1^+$ with a statistical significance of more than 9$\sigma$ over alternative $J^P$ hypotheses ($J^P=0^+,1^-,2^+,2^-$).
The fits yield a mass $(3893.0\pm2.3)~\mevcc$ and a width $(44.2\pm5.4)~\mev$ for the $\zcz$, where the uncertainties are statistical only.
Since the $\zcz$ is also observed in $\ee\to \piz D^*\bar{D}$~\cite{BESIII2015_Zcz} and its mass is close to the mass threshold of $D^*\bar{D}$,
we also perform an alternative fit by parameterizing the $\zcz$ with a Flatt$\acute{e}$ formula as those for $\ee\to\pipi\jpsi$~\cite{2017PingRG}.
The fit results are not sensitive enough to determine the coupling constants $g_{\piz\jpsi}$ and $g_{D^*\bar{D}}$.
However, if the ratio $g_{D^*\bar{D}}$/$g_{\piz\jpsi}$ is fixed to the value reported in Ref.~\cite{2017PingRG}, the fit gives a comparable description of the data as the nominal case.
We try to improve the fit to $M_{\piz\jpsi}$ projections around 3.4 $\gev$ by adding a possible new $Z$ state around 3.4 $\gev$, substituting the $\piz\piz-S$ waves with $K-$matrix approach~\cite{kmatrix}, adding the contribution of direct three-body decays and/or $f_2(1270)$. However, none of these tests make a big difference due to the limited statistics. 

\subsection{STUDY OF $\ee\to\piz\zcz\to\piz\piz\jpsi$ }
Based on the above procedure, the Born cross sections for the process $\ee\to\piz\zcz\to\piz\piz\jpsi$ are measured
using $\sigma^{\rm Born}_{Z_c^0}=f_{ Z_c^0}\times\sigma^{\rm Born}_{\piz\piz\jpsi}$, where $f_{Z_c^0}$ is the fraction of the $\zcz$ component
in the $\ee\to\piz\piz\jpsi$ process, extracted from the PWA.
To obtain the energy-dependent line shape of the cross section for $\ee\to\piz\zcz\to\piz\piz\jpsi$ around the $Y(4220)$, we also perform the PWA for other data samples
with c.m.~energy around the $Y(4220)$, individually, where the properties of the $\zcz$ and its coupling strength to $\piz\jpsi$ are fixed to those obtained
from the simultaneous fit.
The resulting cross sections, shown in Fig.~\ref{fig:zclineshape} and in Table~\ref{tab:crszc}, show a clear structure around 4220~$\mev$.
  \begin{table}[htbp]
  \begin{center}
  \small
  \caption{Summary of the Born cross sections of $\ee\to\piz Z_{c}^{0}\to\piz\piz\jpsi$. The first uncertainties are statistical and the second systematic.}
  \begin{tabular}{c  c}
  \hline\hline
  $\sqrt{s}~(\gev)$  &~~~~~$\sigma_{\ee\to\piz{Z}_{c}^{0}} ~(\rm pb)~~~~~ $            \\ \hline
  4.178    & $1.24\pm0.29 \pm 0.87$      \\
  4.189    & $1.62\pm0.57 \pm 1.13$ \\
  4.200    & $4.65\pm1.06 \pm 3.26$      \\
  4.210    & $2.75\pm0.94 \pm 1.93$      \\
  4.219    & $6.66\pm1.50 \pm 4.68$   \\
  4.226    & $6.14\pm1.09 \pm 4.31$  \\
  4.236    & $7.87\pm1.80 \pm 2.59$     \\
  4.244    & $5.03\pm1.21 \pm 1.38$     \\
  4.258    & $2.34\pm0.82 \pm 0.74$      \\
  4.267    & $1.82\pm0.75 \pm 1.28$      \\
  4.278    & $4.78\pm2.78 \pm 3.36$       \\
  \hline \hline
  \end{tabular}
  \label{tab:crszc}
  \end{center}
  \end{table}

We perform a $\chi^2$ fit to the $\zcz$ cross sections using a coherent sum of a relativistic BW function and $\sigma_{NY}(\sqrt{s})$,
as in Eq.~\ref{eq:expo}, but with a threshold $M_{\rm thd}=M_{\piz}+M_{\zcz}$ for the non-resonant component and a PHSP factor for the two-body decay $R\to\piz\zcz$, which both use the measured mass of $\zcz$ from this analysis.
Two solutions with equal quality are found, as shown in Table~\ref{tab:zclineshape}. The fit curve of one solution is shown in Fig.~\ref{fig:zclineshape}, with a goodness of fit of $\chi^{2}/{\rm ndf}=8.5/5$. The mass $(4231.9\pm5.3)~\mevcc$
and width $(41.2\pm16.0)~\mev$ of the resonant structure are consistent with those of the $Y(4220)$ presented previously.
We also tested the scenarios including the $Y(4320)$ with fixed resonant parameters and/or phase, but none of those improve the fit quality.
Due to the lack of data around 4.3~$\gev$, we cannot rule out a contribution from the $Y(4320)$.

\begin{table}[htbp]
\begin{center}
\begin{small}
\caption{Summary of the fit results to the measured cross sections of $\ee\to\piz\zcz\to\piz\piz\jpsi$. The uncertainties are statistical only.}
\begin{tabular}{c c c }\hline\hline
Parameters  & ~~Solution I~~  & ~~Solution II~~   \\\hline
$p_{0}~(c^2/{\rm MeV})$			 &\multicolumn{2}{c}{ $0.0 \pm 11.3 $}   \\
$p_{1}$			 &\multicolumn{2}{c}{ (1.8$\pm$ 1.9)$\times 10^{-2}$}   \\
$M({R})~(\mevcc)$ 			 	  &\multicolumn{2}{c}{ 4231.9$\pm$5.3 }   \\
$\Gamma_{\rm tot}(R))~(\mev)$        &\multicolumn{2}{c}{ 41.2$\pm$16.0}   \\
$\Gamma_{\rm ee}\BR_{R\to\piz\zcz}~(\rm eV)$   &0.53$\pm$0.15 & 0.22$\pm$0.25    \\
$\phi(R)$ 		&(-103.9$\pm$33.9)$^\circ$      &(112.7$\pm$43.0)$^\circ$      \\
\hline\hline
\end{tabular}
\label{tab:zclineshape}
\end{small}
\end{center}
\end{table}

 \begin{figure}[htbp]
  \centering
  \mbox{
     \includegraphics[width=0.4\textwidth]{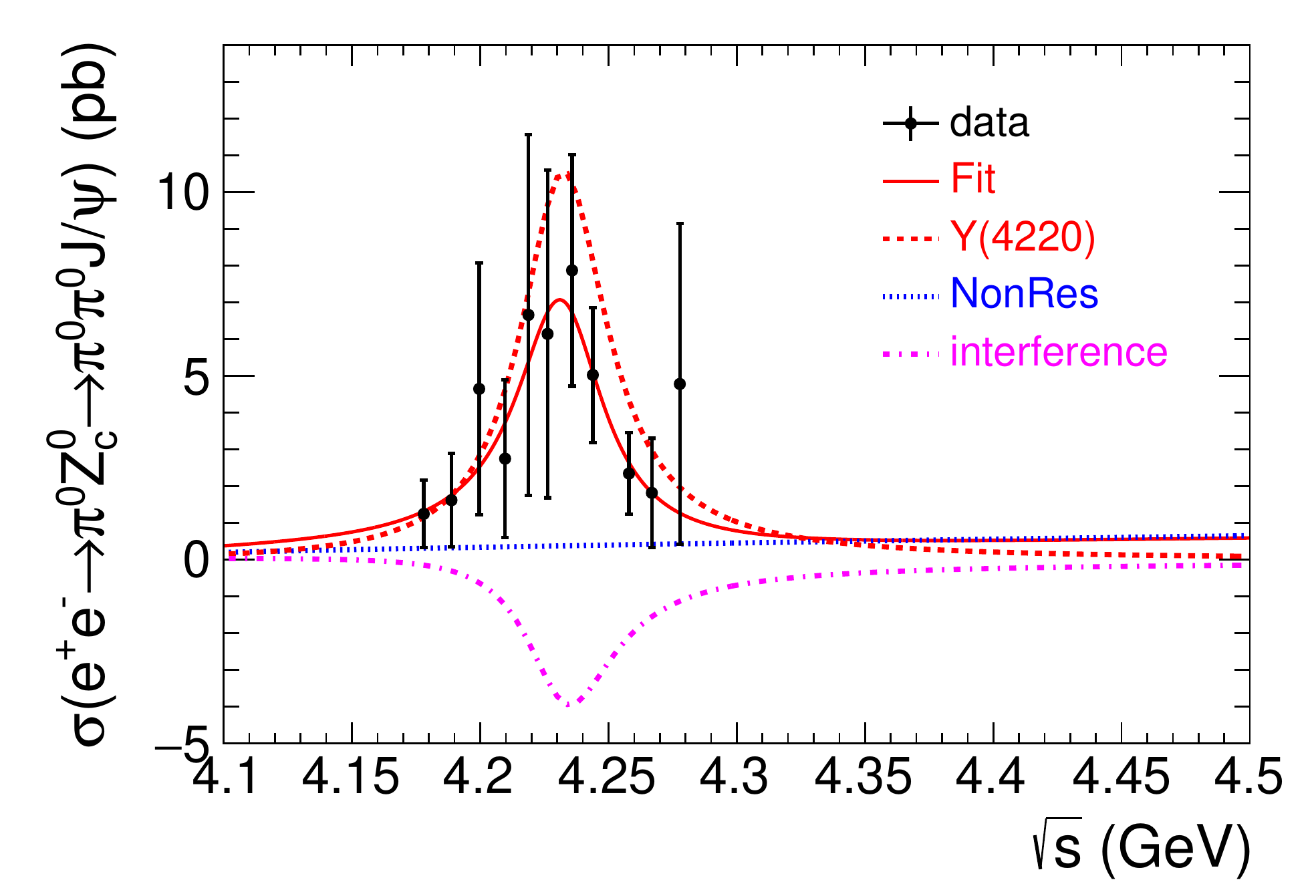}
  }
  \caption{(Color online) Fit to the measured $\ee\to\piz\zcz\to\piz\piz\jpsi$ cross sections. Points with error bars are data, the red solid curve is the total fit result, the red-dashed (blue-dotted) curve is the resonant (non-resonant) component, and the magenta dash-dotted line represents the interference of the two components.}
  \label{fig:zclineshape}
\end{figure}

\subsection{SYSTEMATIC UNCERTAINTIES}

The systematic uncertainties for the $\zcz$ resonant parameters and the corresponding cross sections include those associated with the amplitude modeling and background treatment in the PWA, as well as the detection efficiency difference between data and the MC simulation.

The uncertainties associated with the amplitude modelling in PWA arise from the parameterizations of the intermediate states ($\sigma$, $\fz$, $f_0(1370)$ and $\zcz$),
the radius of angular momentum barrier factor, and possible missing components.
The uncertainties associated with the parameterizations of intermediate states are studied individually by describing $\sigma$ with the PKU ansatz~\cite{E791sigma} or the Zou-Bugg approach~\cite{E791sigma}, varying $f_0(980)$ couplings by 1$\sigma$ of uncertainties~\cite{f0980flatte}, describing $f_0(1370)$ with a mass-dependent width BW function,
and parameterizing $\zcz$ with a Flatt$\acute{e}$-like formula as described for $\ee\to\pipi\jpsi$~\cite{2017PingRG}.
The uncertainty related to the barrier radius $r$ is estimated by varying $r$ in the range 1.0$\sim$5.0~$\gev^{-1}$.
The uncertainty due to extra components is studied by including the process $\ee\to f_2(1270)\jpsi$, which is the most significant (3.2$\sigma$) amplitude not included in the nominal fit.
The uncertainty related to the background treatment is studied by varying the $M_{\LL}$-sideband region.
We perform alternative PWAs for the above scenarios individually and the resultant (largest) changes with respect to the nominal values are regarded as the uncertainties.

The uncertainties related with the detection efficiency include those for tracking of leptons and photons, $\piz$ reconstruction, kinematic fit and ISR correction factor,
as well as the uncertainty by requiring additional MUC hits. 
The uncertainties associated with the previously discussed sources for the measurement of the $\ee\to\piz\piz\jpsi$ cross section are 5.2\%, which directly propagate to the measured cross section of $\zcz$, but do not affect its resonance parameters.
To estimate the uncertainties from the MUC hits requirement, we perform alternative PWA fits by assigning a correction factor for the efficiency to $\jpsi\to\mu^+\mu^-$ MC events according to the $\cos\theta$ distribution of $\mu$, and the resultant changes are regarded as uncertainties.
The detection resolution, about 8.8~$\mevcc$ in the $M_{\piz\jpsi}$ distribution, is not considered in the nominal PWA.
Its effect is estimated with 300 sets of pseudo-experiments. The resultant $|\mu_{pull}|+\sigma_{pull}$ are regarded as the corresponding uncertainties conservatively, where $\mu_{pull}$ and $\sigma_{pull}$ are the mean values and standard deviations of the differences between the input and fitted $\zcz$ values of 300 pseudo-experiments, respectively.
All of the above uncertainties are summarized in Table~\ref{tab:pwasys}. Assuming all of the individuals are uncorrelated, the total uncertainties are the quadratic sum of individual sources.

  \begin{table*}[htbp]
  \begin{center}
  \small
  \caption{Summary of the systematic uncertainties of the $\zcz$ parameters and the cross sections of $\ee\to\piz\zcz\to\piz\piz\jpsi$ in percent ($\%$). }
  \begin{tabular}{l c c c c c c}
      \hline \hline
      \multirow{2}{*}{Sources}
      				&\multirow{2}{*}{~~~$M_{Z_c^0}$~~~}  &\multirow{2}{*}{~~~$\Gamma_{Z_c^0}~~~$}   & \multicolumn{4}{c} {~~~$\piz\zcz$ cross sections~~~ }  \\ \cline{4-7}
				                 &   &     &~~4.226~~ &~~4.236~~ &~~4.244~~ &~~4.258~~ \\ \hline  
   $\zcz$ parametrization  &  $-$    & $-$    &65.7  & 16.5 &11.1 & 12.4   \\
  $\sigma$ parametrization      & 0.03   &4.9 &11.7   & 25.3  & 8.6 & 4.2   \\      
  $f_0(980)$ coupling constant     & 0.01   &0.6  &1.8   & 1.3  & 1.2 & 1.6   \\
  $f_0(1370)$ parametrization      & 0.01   &2.7  &7.5   & 3.2  & 5.6 & 5.6    \\
  $f_2(1270)$ amplitude  	       & 0.05   &2.9 &15.1  & 5.0  & 18.1 & 25.8   \\  
  Barrier radius          		     & 0.01   &13.4  &11.7   & 2.9  & 8.0 & 3.4         \\   
  Background estimation 	       & 0.01   &1.3  &3.8   & 10.2  & 9.8 & 5.1     \\ 
  Event selection 	         & 0.01   &0.2  &5.2   & 5.2  & 5.2 & 5.2     \\
  Detection resolution  	         & 0.06   &11.4  &6.0   & 4.1  & 6.4 & 9.3   \\ \hline
  Total            	               & 0.08   &18.8 &70.4  & 33.3 & 28.0 &32.0        \\
  \hline\hline
  \end{tabular}
  \label{tab:pwasys}
  \end{center}
  \end{table*}
  
The uncertainties for the resonant parameters in the fit to $\sigma(\ee\to\piz\zcz\to\piz\piz\jpsi)$, shown in Table~VIII, include the sources discussed in the fit to $\sigma(\ee\to\piz\piz\jpsi)$ since the same approach is used. 
In this case, the uncertainty of the fit procedure is estimated by replacing the nominal PDF model with one BW function only and by changing the
constant full width of BW to a
phase-space dependent width, $\Gamma\frac{\Phi(\sqrt{s})}{\Phi(M)}$.
The total systematic uncertainties are $4.9~\mevcc$ and $16.4~\mev$ for the mass and width of the structure, respectively.

  \begin{table}[htbp]
  \begin{center}
  \small
  \caption{Summary of the systematic uncertainties of the structure parameters observed in the $\sigma(\ee\to\piz\zcz\to\piz\piz\jpsi)$ line shape. }
  \begin{tabular}{ l c c }
  \hline \hline
  \multirow{2}{*}{Sources} 	& \multicolumn{2}{c}{Uncertainties}  \\ \cline{2-3}
				& Mass ($\mevcc$)  &$\Gamma$($\mev$)  \\ \hline
      Cross section measurement &0.6   & 13.6          \\
      c.m. energies    &0.8   & $-$          \\
      c.m. energy spread  &0.01   & 0.4          \\
      Fit procedure &4.7   & 9.2     \\ \hline
      Total        &4.9   & 16.4           \\
  \hline\hline
  \end{tabular}
  \end{center}
  \label{tab:zclineshape_sys}
  \end{table}

\section{SUMMARY}
In summary, we measured the Born cross sections of $\ee\to\piz\piz\jpsi$ for c.m.~energies between 3.808 and 4.600~$\gev$ with data samples collected by the BESIII experiment.
The measured cross sections are fitted by including two resonant structures, $Y(4220)$ and $Y(4320)$, with the resonant
parameters of the $Y(4320)$ fixed to the values taken from Ref.~\cite{LiuZQ2017}.
The mass and width of the $Y(4220)$ are measured to be  $(4220.4\pm2.4\pm2.3)~\mevcc$ and $(46.2\pm4.7\pm2.1)~\mev$, respectively, where the first uncertainties are statistical, and the second are systematic (the same as following). These measurements agree with those reported in Ref.~\cite{LiuZQ2017}, and confirm the existence of the $Y(4220)$.
The average ratio of the cross section $\ee\to\piz\piz\jpsi$ to that of $\ee\to\pip\pim\jpsi$~\cite{LiuZQ2017} is $0.48\pm0.02$, which is consistent with isospin symmetry.

The $\zcz$ signal is clearly observed in the $M_{\piz\jpsi}$ distribution, and a PWA is performed to study its properties. The spin-parity of the $\zcz$ is
determined to be $J^{P}=1^+$, and the measured mass $(3893.0\pm2.3\pm19.9)~\mevcc$ and width $(44.2\pm5.4\pm9.1)~\mev$ correspond to a pole position $(3893.1\pm2.2\pm3.0)-i(22.2\pm2.6\pm7.0)~\mevcc$, which is the complex zero of the denominator of the BW. 
These values are consistent with those of the
charged $Z_c(3900)^\pm$ observed in $\ee\to\pipi\jpsi$.
The Born cross sections of $\ee\to\piz\zcz\to\piz\piz\jpsi$ are also measured and fitted for c.m.~energies between 4.178 and 4.278~$\gev$.
The fit yields a structure with a mass of $(4231.9\pm5.3\pm4.9)~\mevcc$ and a width of $(41.2\pm16.0\pm16.4)~\mev$, compatible with the $Y(4220)$.
The relationship between the two exotic states $Y(4220)$ and the $\zcz$ is established for the first time.
Due to the lack of data around 4.3~GeV, the existence of the $Y(4320)$ in the $\zcz$ production cannot be ruled out.

\section{ACKNOWLEDGMENTS}
The BESIII collaboration thanks the staff of BEPCII and the IHEP computing center and the supercomputing center of USTC for their strong support. This work is supported in part by National Key Basic Research Program of China under Contract No. 2015CB856700; National Natural Science Foundation of China (NSFC) under Contracts Nos. 11335008, 11375170, 11475164, 11475169, 11625523, 11605196, 11605198, 11705192; the Chinese Academy of Sciences (CAS) Large-Scale Scientific Facility Program; 
Joint Large-Scale Scientific Facility Funds of the NSFC and CAS under Contracts Nos.  U1532102, U1732263, U1832103; CAS under Contracts Nos. QYZDJSSW-SLH003, QYZDJ-SSW-SLH040; 100 Talents Program of CAS; INPAC and Shanghai Key Laboratory for Particle Physics and Cosmology; ERC under Contract No. 758462; German Research Foundation DFG under Contract Nos. Collaborative Research Center CRC-1044, ROF 2359; Istituto Nazionale di Fisica Nucleare, Italy; Ministry of Development of Turkey under Contract No. DPT2006K-120470; National Science and Technology fund; STFC (United Kingdom); The Knut and Alice Wallenberg Foundation (Sweden) under Contracts Nos. DH140054, DH1600214; The Swedish Research Council; U. S. Department of Energy under Contracts Nos. DE-FG02-05ER41374, DE-SC-0010118, DE-SC-0012069; University of Groningen (RuG) and the Helmholtzzentrum fuer Schwerionenforschung GmbH (GSI), Darmstadt.

\end{document}